\documentclass[aip,apl,twocolumn,preprintnumbers,amsmath,amssymb]{revtex4}

\usepackage{graphicx}
\usepackage[colorlinks=true,citecolor=blue]{hyperref}

\newcommand{\D}[2]{\frac{\mbox{d} {#1}}{\mbox{d}{#2}}}
\def\bra#1{\mathinner{\langle{#1}|}}
\def\ket#1{\mathinner{|{#1}\rangle}}
\def\braket#1{\mathinner{\langle{#1}\rangle}}

\begin{document}
    \title{Nonlinear Temporal Dynamics of Strongly Coupled Quantum Dot-Cavity System}
    \author{Arka Majumdar}
    \email{arkam@stanford.edu}
    \author{Dirk Englund$^\dag$}
    \author{Michal Bajcsy}
    \author{Jelena Vu\v{c}kovi\'{c}}
    \affiliation{E.L.Ginzton Laboratory, Stanford University, Stanford, CA, $94305$\\
    $^\dag$ Department of Electrical Engineering and Department of Applied Physics, Columbia University, New York, NY $10027$}

\begin{abstract}
We theoretically analyze and simulate the temporal dynamics of
strongly coupled quantum dot-cavity system driven by a resonant
laser pulse. We observe the signature of Rabi oscillation in the
time resolved response of the system (i.e., in the numerically
calculated cavity output), derive simplified linear and non-linear
semi-classical models that approximate well the system's behavior
in the limits of high and low power drive pulse, and describe the
role of quantum coherence in the exact dynamics of the system.
Finally, we also present experimental data showing the signature
of the Rabi oscillation in time domain.
\end{abstract}
\maketitle
A single quantum dot (QD) coupled to a photonic crystal
micro-cavity constitutes an integrated nano-photonic platform for
probing solid state cavity quantum electrodynamic (QED) effects
\cite{andrei_njp}. The eigenstates of this coupled system form an
anharmonic ladder, which results in an optical nonlinearity at a
single photon level. In recent years, this nonlinearity has been
used to perform all-optical \cite{arka_switching,edo_switching}
and electro-optic switching \cite{andrei_eom} as well as to
generate non-classical states of photons \cite{article:faraon08,
AM_tunneling, Imma_blockade}.

\begin{figure}
\centering
\includegraphics[width=3.5in]{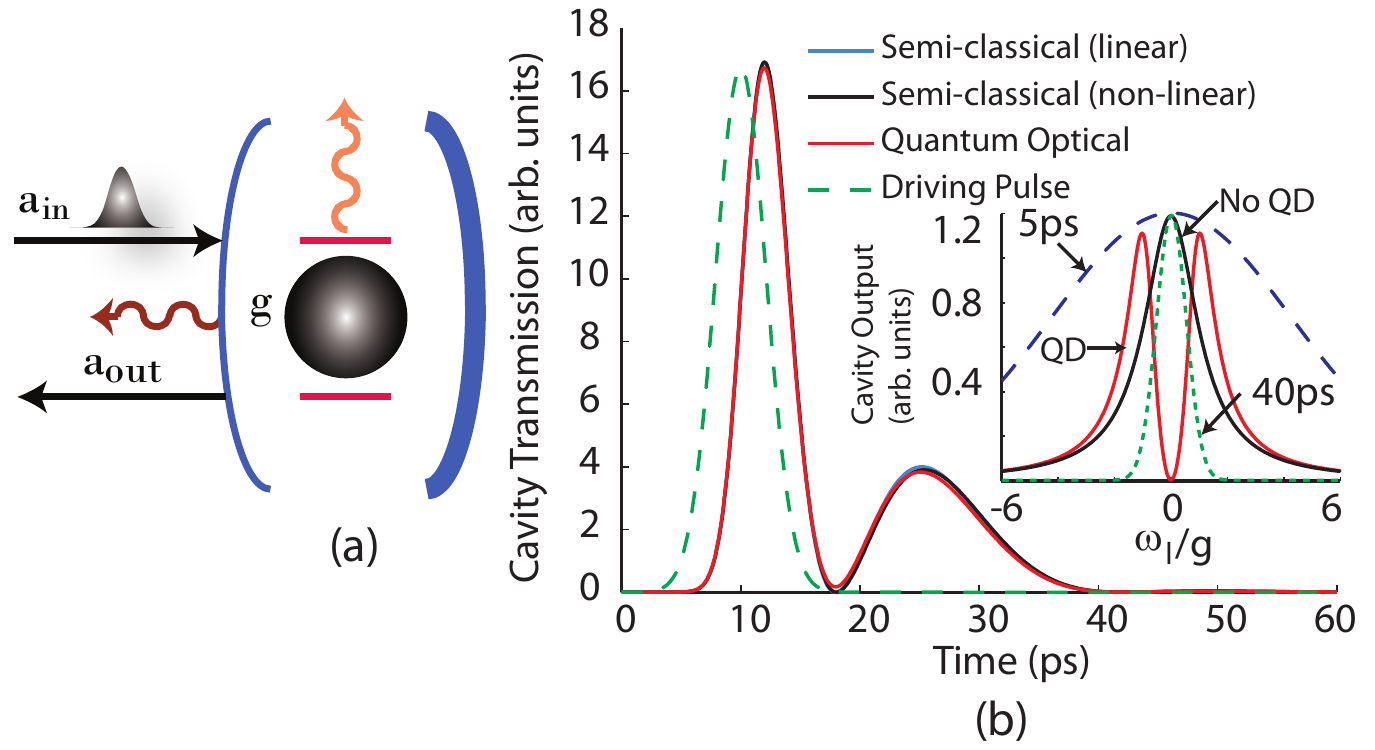}
\caption{(color online) (a) The schematic of the coupled QD-cavity
system. It is driven by a laser pulse, and the cavity output is
monitored. (b) The cavity transmission calculated by three
different models: the quantum optical (red), semi-classical linear
(blue) and non-linear (black) model at low ($\Omega_0/2\pi=1$ GHz)
peak intensity of the driving pulse. All three models match quite
well. The input pulse is also shown (green dashed line). The
oscillation in the cavity output is due to Rabi oscillation of the
photon between the QD and the cavity. Inset shows the cavity
transmission spectrum in presence and in absence of the strongly
coupled QD. The split resonances are separated approximately by
twice the coherent dot-cavity interaction strength $g$. The
spectral shape of laser pulses with pulse-length $5$ ps (blue
dashed line) and $40$ ps (green dashed line) is also shown.
Parameters used for the simulations are $g/2\pi=25$ GHz,
$\kappa/2\pi=29$ GHz and $\gamma/2\pi=1$ GHz.} \label{fig1_intro}
\end{figure}

In this paper, we study the temporal dynamics of the  coupled
dot-cavity system driven by a short laser pulse (Fig.
\ref{fig1_intro} a) using a full quantum optical numerical
simulation. The oscillatory behavior of the cavity output (Fig.
\ref{fig1_intro} b), which is caused by the vacuum Rabi splitting,
is analyzed at low, intermediate, and high intensity of the
driving laser.  Specifically, we derive a linear semi-classical
description of the system, and show that under weak driving, the
coupled QD-cavity system follows the same dynamics as a set of two
classical linear coupled oscillators. Following this, we describe
an improved, non-linear semi-classical model, that mimics the
quantum optical model very well for both very low and high peak
intensity of the driving pulse. However, the non-linear
semi-classical model deviates from quantum optical description at
intermediate peak intensities of the drive pulse and we show that
this discrepancy arises from the coherence present in the quantum
optical system. Finally, we present a study of the temporal
dynamics as a function of the major parameters describing the
cavity-QD system as well as experimental data showing the
signature of the Rabi oscillation in time domain.
Under rotating wave approximation, the quantum-mechanical
Hamiltonian $\mathcal{H}$ describing the coherent dynamics of the
coupled system is given by
\begin{equation}
\mathcal{H}=\omega_a\sigma^\dag\sigma+\omega_c a^\dag a+
ig(a\sigma^\dag-a^\dag\sigma).
\end{equation}
Here, $\omega_c$ and $\omega_a$ are, respectively, the resonance
frequencies of the cavity and the QD; $a$ is the annihilation
operator for the cavity mode; $\sigma=|g\rangle\langle e|$ is the
lowering operator for the QD with excited state $|e\rangle$ and
ground state $|g\rangle$; $g$ is the coherent interaction strength
between the QD and the cavit and $\hbar$ is set to $1$. When this
system is coherently driven by a laser pulse with strength
$\Omega(t)=\Omega_0p(t)$ and a center frequency $\omega_l$, the
driven Hamiltonian in a frame rotating at the frequency $\omega_l$
is
\begin{equation}
H=\Delta_ca^\dag a+\Delta_a\sigma^\dag \sigma+ig(a^\dag\sigma-
a\sigma^\dag )+i\Omega(t)(a-a^\dag).
\end{equation}
Here,  $\Delta_c$ and $\Delta_a$ are the detuning of the cavity
and the QD resonance from the laser frequency; $\Omega_0$ is the
maximum laser strength and $p(t)$ is proportional to the envelope
of the laser electric field. The dynamics of the lossy system is
determined by using the Master equation
\begin{equation}
\label{Maseq} \frac{d\rho}{dt}=-i[H,\rho]+ 2\kappa
\mathcal{L}[a]+2\gamma \mathcal{L}[\sigma]
\end{equation}
where $\rho$ is the density matrix of the coupled QD-cavity
system; $\kappa$ is the cavity field decay rate and $\gamma$ is
the dipole spontaneous emission rate. $\mathcal{L}[D]$ is the
Lindblad operator corresponding to a collapse operator $D$. This
is used to model the incoherent decays and is given by:
\begin{equation}
\mathcal{L}[D]= D\rho D^\dag-\frac{1}{2}D^\dag D
\rho-\frac{1}{2}\rho D^\dag D
\end{equation}
The Master equation is solved using numerical integration routines
provided in quantum optics toolbox, truncating the photon states
to $20$ photons \cite{qotoolbox}. This method is completely
quantum mechanical, and no approximation (other than the standard
Born-Markov approximation and truncation of Fock state basis) is
made.

A semi-classical description of the coupled system \cite{Arka_eom} can be derived by
using the relation
\begin{equation}
\frac{d\langle D \rangle}{dt}=Tr\left[D\frac{d\rho}{dt}\right]
\end{equation}
valid for any operator $D$. The mean field dynamical equations for the
coupled QD-cavity system can then be written as
\begin{eqnarray}
\label{mean_field_eq1}
\D{\langle a \rangle}{ t } &=& -\kappa \braket{a}+g \braket{\sigma}- \sqrt{\kappa} \Omega(t) \\
\label{mean_field_eq2}
\D{\langle \sigma \rangle}{ t } &=& -\gamma \braket{\sigma}+g \braket{a \sigma_z}\\
\label{mean_field_eq3} \D{\langle \sigma_z \rangle}{ t }
&=&-2\gamma(\braket{\sigma_z}+1)-2g(\braket{a^\dag\sigma}+\braket{a\sigma^\dag})
\end{eqnarray}
where, $\sigma_z=\ket{e}\bra{e}-\ket{g}\bra{g}$. We note that this
set of equations is not complete and to exactly solve this, we
need to find the equations describing all the other higher order
moments, namely $\braket{a \sigma_z}$ and $\braket{a
\sigma^\dag}$. However, in the low excitation regime (no more than
$1$ photon in the system) the QD will remain mostly in its ground
state and we can approximate $\braket{\sigma_z}\approx-1$ and
replace $\braket{a \sigma_z}=-\braket{a}$. The resulting set of
equations
\begin{eqnarray}
\D{\langle a \rangle}{ t } &=& -\kappa \braket{a}+g \braket{\sigma}- \sqrt{\kappa} \Omega(t) \\
\D{\langle \sigma \rangle}{ t } &=& -\gamma \braket{\sigma}-g\braket{a} \label{linear2}
\end{eqnarray}
is identical to the set of equations describing the dynamics of
two coupled linear classical oscillators (see Appendix \ref{app}).
Although this approximation neglects the nonlinear nature of the
QD, it matches the actual output quantitatively at low excitation
power. Unfortunately, with increasing drive intensities this model
fails completely, as the approximation
$\braket{\sigma_z}\approx-1$ becomes invalid. For sufficiently
high drive intensities though $\braket{\sigma_z}\to0$ and equation
(\ref{linear2}) simplifies to
\begin{equation}
\D{\langle \sigma \rangle}{ t } = -\gamma \braket{\sigma}
\label{linear3}
\end{equation}

Alternatively, we can retain the dynamics of the $\sigma_z$, while
making the set of equations (\ref{mean_field_eq1}),
(\ref{mean_field_eq2}), and (\ref{mean_field_eq3}) complete by
using the approximations
$\braket{a\sigma_z}\approx\braket{a}\braket{\sigma_z}$ and
$\braket{a^\dag\sigma}\approx\braket{a^\dag}\braket{\sigma}$
\cite{armen_hideo_PRA}. While this approach neglects the coherence
of the system while analyzing the mean-field dynamical equations,
the nonlinear behavior of the QD is taken into account.

The temporal cavity outputs at low excitation power match very
well for the three different models (Fig. \ref{fig1_intro}b). For
the numerical simulation, we used a Gaussian pulse with full-width
half-maximum (FWHM) of $5$ ps as we want to drive the dot-cavity
system with a pulse having bandwidth higher than the coupled
system (as shown in the inset of Fig. \ref{fig1_intro}b). An
oscillation in the cavity output is observed. This oscillation is
due to the coherent Rabi oscillation of the photons between the QD
and the cavity. Note that further oscillations are quenched by the
decay of the cavity field.

To intuitively understand the origin of the oscillation, we
analytically solve the linear semi-classical equations to find the
eigenvalues of the lossy coupled system as
\begin{equation}
\omega_{\pm}=\frac{\omega_c+\omega_d}{2}-i\frac{\kappa+\gamma}{2}
\pm \sqrt{g^2+\frac{1}{4}\left(\delta-i(\kappa-\gamma)\right)^2}
\end{equation}
where $\delta=\Delta_c-\Delta_a$ is the dot-cavity detuning. When
the real part of the expression under the square root is positive,
the system is in strong coupling regime and a split resonance
appears in the cavity transmission spectrum (inset of Fig.
\ref{fig1_intro} b). Without a coupled QD, a single Lorentzian
peak is observed in the cavity transmission. The two peaks are
entangled states of the QD and the cavity, known as polaritons.
When the cavity is driven with a short pulse, having bandwidth
more than the coupled system, the cavity output is modulated at
the frequency difference between the polaritons. i.e.,
$2\sqrt{g^2+\frac{1}{4}\left(\delta-i(\kappa-\gamma)\right)^2}$.

\begin{figure}
\centering
\includegraphics[width=3.25in]{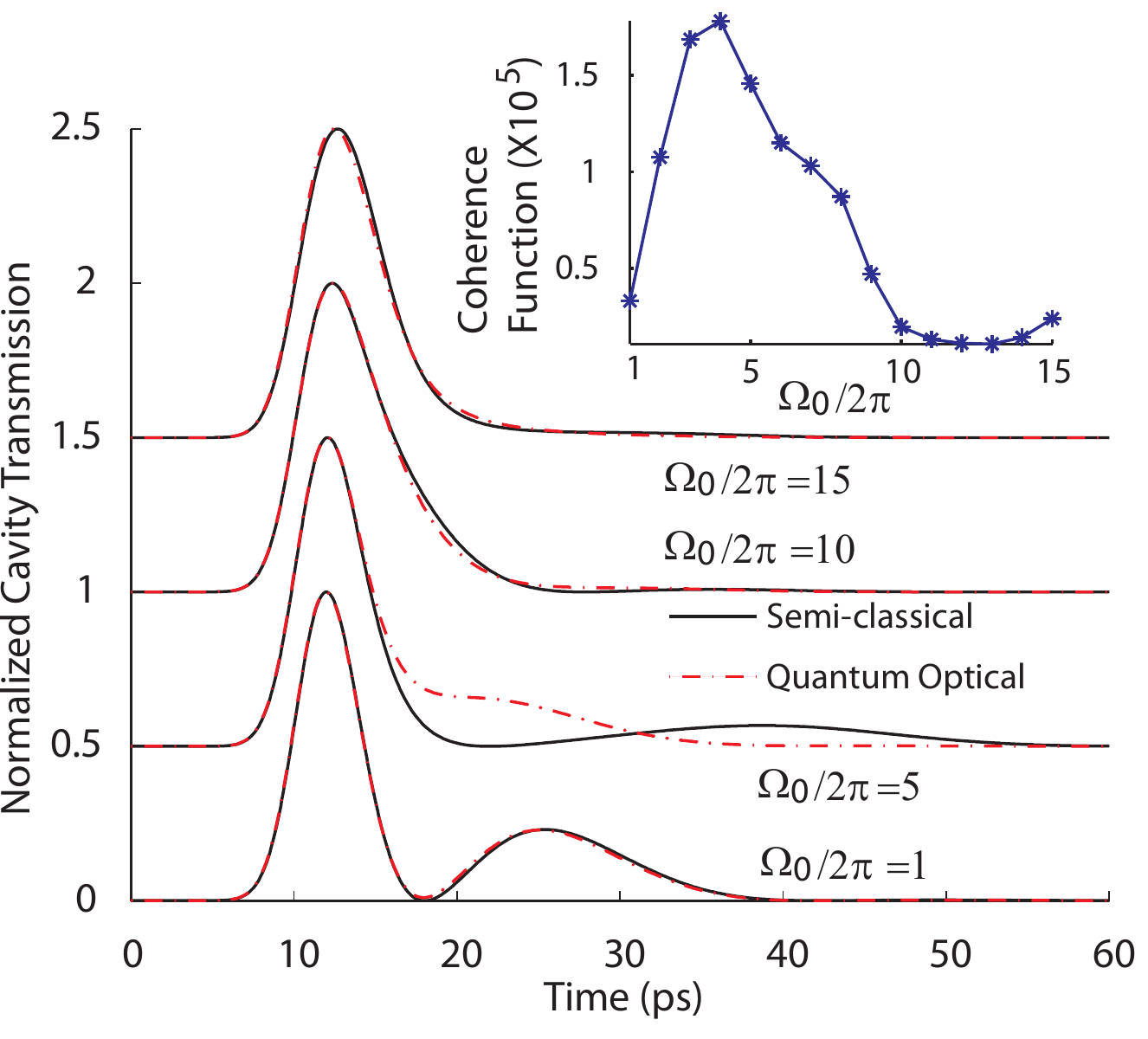}
\caption{(color online) (a) Comparison between the temporal cavity
transmission obtained via quantum optical (red dashed line) and
the semi-classical non-linear (black solid line) models. The
cavity transmission is normalized by the maximum cavity
transmission and plots are vertically offset for clarity. The two
models match quite well at low and high driving power, but at
intermediate power, they differ. Inset shows the coherence
calculated as $(\langle a^\dag\sigma\rangle-\langle
a^\dag\rangle\langle\sigma\rangle)/\Omega_0^2$ integrated over
time as a function of the driving strength $\Omega_0$. We observe
that quantity increases in the intermediate driving power.
Parameters used for the simulations: $g/2\pi=25$ GHz and
$\kappa/2\pi=29$ GHz.} \label{fig2_nl_qo}
\end{figure}

Although the non-linear semi-classical model allows QD saturation,
it neglects the quantum mechanical coherence between the QD and
the cavity. Fig. \ref{fig2_nl_qo} compares the semiclassical and
quantum optical simulations of the coupled dot-cavity system. We
find that the results match well both at low (when the QD excited
state population is almost zero and $\sigma_z\sim -1$) and high
intensities of the drive (when the QD is saturated and
$\sigma_z\sim0$). As expected, the nonlinear semi-classical
approach will deviate for intermediate intensities. We plot the
quantity $(\langle a^\dag\sigma\rangle-\langle
a^\dag\rangle\langle\sigma\rangle)/\Omega_0^2$ integrated over
time as a function of the driving strength $\Omega_0$ in the inset
of Fig. \ref{fig2_nl_qo}. This quantity is zero in absence of any
coherence. We observe that this quantity is smaller for both low
and high excitation power, compared to the value at intermediate
excitations. Note that the onset of increase in the higher
excitation power is due to numerical errors caused by the
truncated Fock state basis.

\begin{figure}
\centering
\includegraphics[width=3.5in]{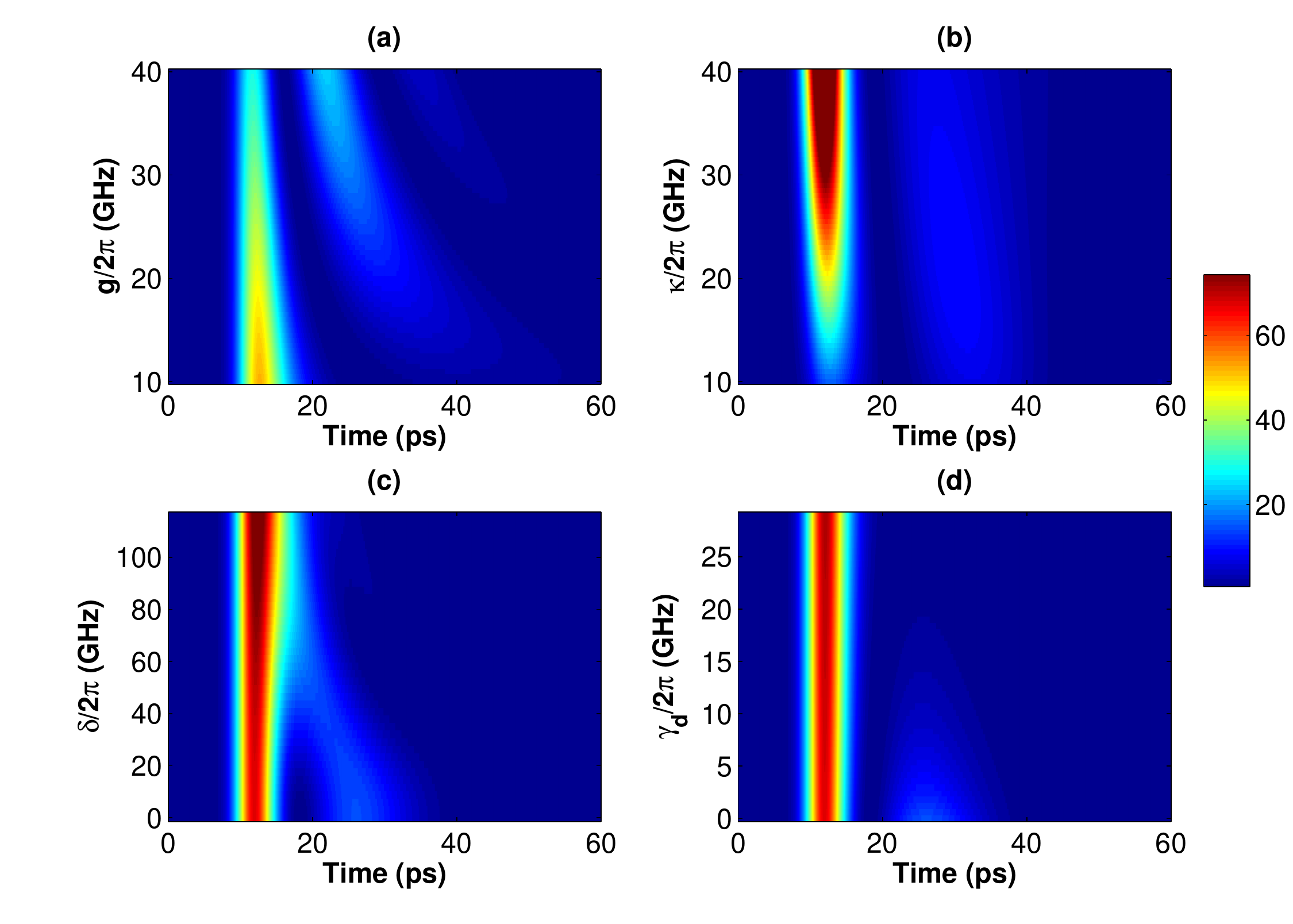}
\caption{(color online) The temporal cavity output obtained from
the full quantum optical simulation as a function of (a) the
dot-cavity coupling strength $g$ (here $\kappa/2\pi=20$ GHz;
$\delta=0$ and $\gamma_d=0$), (b) the cavity field decay rate
$\kappa$ (here $g/2\pi=20$ GHz; $\delta=0$ and $\gamma_d=0$), (c)
the dot cavity detuning $\delta$ (here $g/2\pi=\kappa/2\pi=20$ GHz
and $\gamma_d=0$) and (d) the pure QD dephasing rate $\gamma_d$
(here $g/2\pi=\kappa/2\pi=20$ GHz and $\delta=0$). For all the
simulations a low excitation power ($\Omega_0/2\pi=2$) is
assumed.} \label{fig3_param}
\end{figure}

Finally, we analyze the dependence of the temporal cavity output
as a function of four quantities: dot-cavity detuning $\delta$,
the dot-cavity coupling rate $g$, the cavity field decay rate
$\kappa$ and pure QD dephasing rate $\gamma_d$. We observe an
increase in oscillation frequency (decrease in the time interval
between the two peaks) when we increase $g$ (Fig.
\ref{fig3_param}a). This is consistent with the oscillation period
as predicted by the simple linear analysis. At the same time, the
oscillation period depends only weakly on $\kappa$ (Fig.
\ref{fig3_param}b). We note that an increasing cavity output with
increasing cavity decay rate $\kappa$. This is due to the
increasing overlap between the input pulse and the cavity
spectrum. The oscillation frequency increases with increasing
detuning between the dot and the cavity and when the QD is detuned
too far from the cavity, the oscillation almost disappears. This
is expected, as with large enough detuning the input pulse is not
affected by the QD (Fig. \ref{fig3_param}c). An important quantity
in solid-state cavity QED is pure QD dephasing, which destroys the
coherence of the system, without affecting any population of the
quantum dot states.The effect of pure QD dephasing can be
incorporated by adding a term
$2\gamma_d\mathcal{L}(\sigma^\dag\sigma)$ in the Master equation
\cite{majumdar_phonon_11}, where $\gamma_d$ is the pure QD
dephasing rate. Fig. \ref{fig3_param}d shows the cavity output as
a function of pure QD dephasing rate $\gamma_d$ and we observe
that the oscillation eventually disappears with increasing
dephasing.

\begin{figure}
\centering
\includegraphics[width=3.5in]{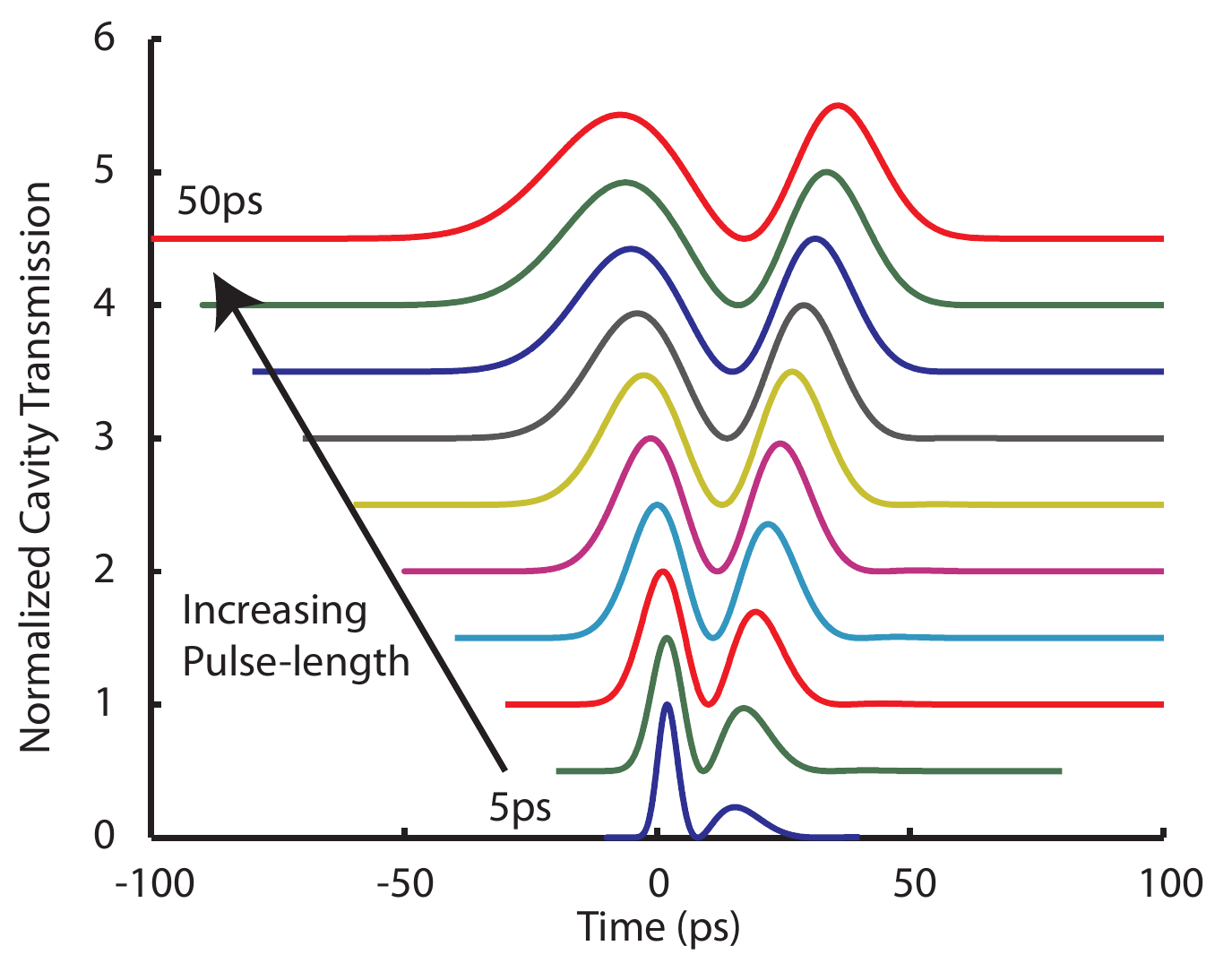}
\caption{(color online) The normalized cavity transmission for
different pulse duration. The pulse duration is changed from $5$
ps to $50$ ps. We observe oscillation in the cavity output,
although the oscillation frequency decreases with increasing
pulse-width. This can be explained by the reduced overlap between
the pulse and the coupled dot-cavity system in frequency domain.}
\label{fig4_th_pulse}
\end{figure}

\begin{figure*}
\centering
\includegraphics[width=6in]{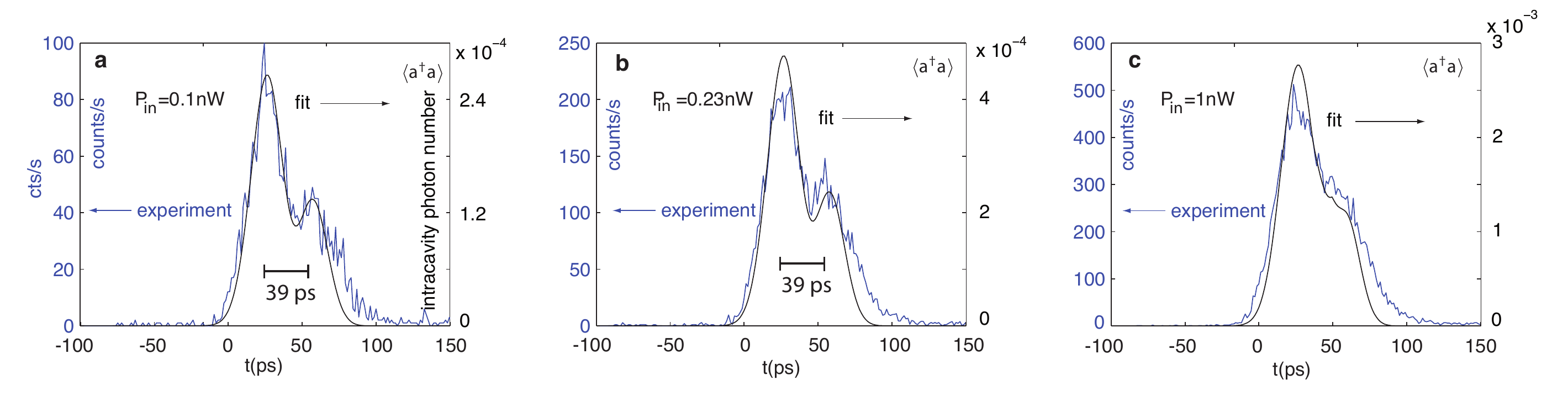}
\caption{(color online) Experimentally measured time resolved
transmission of $40$ps pulses through a strongly coupled
dot-cavity system for three different powers (averaged over the
pulse repetition period): (a)$0.1$ nW; (b) $0.23$ nW; (c) $1$ nW.
The powers are measured in front of the objective lens in the
confocal microscopy setup. For this specific system cavity field
decay rate $\kappa/2\pi=29$ GHz and coherent dot-cavity coupling
strength $g/2\pi=25$ GHz. Clear oscillations are observed in the
cavity transmission, consistent with the the theoretical
predictions. We also observe decreasing oscillation with
increasing laser power due to QD saturation.}\label{fig4_exp}
\end{figure*}

Finally, we analyze the dependence of the cavity transmission on
the pulse duration (Fig. \ref{fig4_th_pulse}). The pulse duration
is changed from $5$ ps to $50$ ps. We observe that oscillation
frequency is decreasing with increasing pulse length. This can be
explained by the reduced overlap between the input pulse and the
coupled dot-cavity system with reduction in pulse bandwidth. In
other words, a long pulse does not have sufficient bandwidth to
excite both the polaritons (as shown in the inset of Fig.
\ref{fig1_intro} b) and the oscillation frequency in the cavity
output deviates more from $2$g.
To test the validity of our numerical simulations, we
experimentally probed a strongly coupled QD-cavity system. A
cross-polarized reflectivity setup was used to obtain the
transmission of light through the coupled system and the cavity
transmission was monitored with a Hamamatsu streak camera. Details
of the fabrication and the experimental setup can be found in Ref.
\cite{article:eng07} with the experimental parameters of the
probed dot-cavity system being $g/2\pi=25$ GHz and
$\kappa/2\pi=29$ GHz \cite{arka_switching}. Unfortunately, we did
not observe the predicted oscillations in the initial experiments
measuring the transmission of $5$ ps pulses through the cavity.
This was  most likely caused by the limited time resolution of our
detector. Subsequently the experiment was performed with a longer
pulse ($40$ ps FWHM). A long pulse does not have sufficient
bandwidth to excite both the polaritons (as shown in the inset of
Fig. \ref{fig1_intro} b) and the oscillation frequency in the
cavity transmission is different from the value $~2g$, as shown in
Fig. \ref{fig4_th_pulse}. Fig. \ref{fig4_exp} a,b,c show the
experimentally obtained cavity output for three different
excitation powers. The experimental data match qualitatively the
predictions from the numerical simulation and clear oscillation is
observed in the cavity output. This oscillation disappears with
increasing laser power, as expected from the QD saturation. The
oscillation period is estimated to be $~25$ GHz, corresponding to
time difference of $39$ ps between the two peaks. We note that the
numerically obtained plots in Fig. \ref{fig4_th_pulse} are done
with very small excitation power. However, the experiment cannot
be performed with such low excitation power as the detected signal
is too low. Hence, in the experiment, the coupled system is driven
close to the QD saturation, and the oscillations are much less
visible.

In summary, we have analyzed the nonlinear temporal dynamics of a
strongly coupled QD-cavity system driven by a short laser pulse.
We showed that this quantum optical system behaves similar to two
coupled classical linear oscillators when the system is driven
with a weak pulse and that a signature of the vacuum Rabi
oscillations can be observed in the time resolved cavity
transmission. For high intensity pulse these oscillations die down
due to saturation of the QD. We provided a semi-classical
non-linear model and showed that in the actual dynamics the role
of quantum coherence is important. Lastly, we presented
experimental evidence of those oscillations in the cavity output.

Financial support was provided by the Office of Naval Research
(PECASE Award), National Science Foundation, and Army Research
Office. A.M. was supported by the SGF (Texas Instruments Fellow).
Work was performed in part at the Stanford Nanofabrication
Facility of NNIN supported by the National Science Foundation.
D.E. acknowledges support by the Sloan research fellowship and by
the U.S. Air Force Office of Scientific Research Young
Investigator Program, AFOSR Grant No. FA$9550-11-1-0014$, supervised
by Dr. Gernot Pomrenke.

\begin{appendix}
\section{Dynamics of two classical coupled linear oscillators}
\label{app}

The dynamics of two classical coupled oscillators, with resonance frequency $\omega_0$ and decay rates $\Gamma_1$ and $\Gamma_2$, are governed by
\begin{equation}
\frac{d^2x_1}{dt^2}+\Gamma_1\frac{dx_1}{dt}+\omega_0^2x_1+G(x_1-x_2)=\Omega(t)e^{i\omega_0t}
\end{equation}
and
\begin{equation}
\frac{d^2x_2}{dt^2}+\Gamma_2\frac{dx_2}{dt}+\omega_0^2x_2+G(x_2-x_1)=0,
\end{equation}
where $G$ denotes the coupling strength between the oscillators.
One of the oscillators is driven resonantly with driving strength $\Omega(t)$, as the cavity is driven by a laser. We assume the solution of the form $x_1(t)=X_1(t) e^{i\omega_0t}$
and $x_2(t)=X_2(t)e^{i\omega_0t}$, where $X_1(t)$ and $X_2(t)$ are
slowly varying envelopes of the actual oscillator outputs. Then we can write
\begin{eqnarray*}
  \frac{dx_1}{dt} &=& i\omega_0X_1 e^{i\omega_0t}+\left(\frac{dX_1}{dt}e^{i\omega_0t}\right)\\
  \frac{d^2x_1}{dt^2} &=& 2i\omega_0\frac{dX_1}{dt}e^{i\omega_0t}-\omega_0^2X_1e^{i\omega_0t}+\left(\frac{d^2X_1}{dt^2}e^{i\omega_0t}\right)
\end{eqnarray*}
For $x_2$ we can find similar equations. Using slowly varying
envelope approximation ($\frac{dX_1}{dt}<<i\omega_0X_1$ and $\frac{d^2X_1}{dt^2}<<i\omega_0\frac{dX_1}{dt}, \omega_0^2X_1$),
we remove the bracketed terms and get the
following equations for the un-driven coupled oscillator system:
\begin{equation}
\frac{dX_1}{dt}=-\left(\frac{\Gamma_1}{2}+\frac{G}{2i\omega_0}\right)X_1+\frac{G}{2i\omega_0}X_2+\Omega(t)
\end{equation}
and
\begin{equation}
\frac{dX_2}{dt}=-\left(\frac{\Gamma_2}{2}+\frac{G}{2i\omega_0}\right)X_2+\frac{G}{2i\omega_0}X_1
\end{equation}
\end{appendix}

\bibliography{Pulsed_QDPC}
\end{document}